\documentclass[lettersize,journal]{IEEEtran}
\IEEEoverridecommandlockouts
\usepackage{cite}
\usepackage{amsmath,amssymb,amsfonts}
\usepackage{algorithmic}
\usepackage{graphicx}
\usepackage{textcomp}
\usepackage{xcolor}
\usepackage{comment}
\usepackage{booktabs}
\def\BibTeX{{\rm B\kern-.05em{\sc i\kern-.025em b}\kern-.08em
    T\kern-.1667em\lower.7ex\hbox{E}\kern-.125emX}}
\begin{document}

\title{SW-TNC : Reaching the Most Complex Random Quantum Circuit via Tensor Network Contraction
}

\author{
\IEEEauthorblockN{
Yaojian Chen\textsuperscript{1}, Zhaoqi Sun\textsuperscript{2}, Chengyu Qiu\textsuperscript{1}, Zegang Li\textsuperscript{1}, Yanfei Liu\textsuperscript{3}, \\
Lin Gan\textsuperscript{1,3,*}, Xiaohui Duan\textsuperscript{4} and Guangwen Yang\textsuperscript{1,3} \thanks{*Corresponding Author, Lin Gan: \underline{lingan@tsinghua.edu.cn}
}}

\IEEEauthorblockA{
\textsuperscript{1}Tsinghua University, China
}
\IEEEauthorblockA{
\textsuperscript{2}Zhengzhou University, China
}
\IEEEauthorblockA{
\textsuperscript{3}National Supercomputing Center in Wuxi, China
}
\IEEEauthorblockA{
\textsuperscript{4}Shandong University, China
}
}

\maketitle

\begin{abstract}
Classical simulation is essential in quantum algorithm development and quantum device verification. With the increasing complexity and diversity of quantum circuit structures, existing classical simulation algorithms need to be improved and extended. In this work, we propose novel strategies for tensor network contraction based simulator on Sunway architecture. Our approach addresses three main aspects: complexity, computational paradigms and fine-grained optimization. Data reuse schemes are designed to reduce floating-point operations, and memory organization techniques are employed to eliminate slicing overhead while maintaining parallelism. Step fusion strategy is extended by multi-core cooperation to improve the data locality and computation intensity. Fine-grained optimizations, such as in-kernel vectorized permutations, and split-K operators, are developed as well to address the challenges in new hotspot distribution and topological structure. These innovations can accelerate the simulation of the Zuchongzhi-60-24 by more than 10 times, using more than 1024 Sunway nodes (399,360 cores). Our work demonstrates the potential for enabling efficient classical simulation of increasingly complex quantum circuits.
\end{abstract}


\section{Introduction}
In the NISQ era\cite{Preskill_2018}, classical simulation of quantum circuits (i.e. quantum circuit simulation) is essential to verify quantum hardware and algorithms, as quantum hardware still suffers from extremely low fidelity as $~0.2\%$ \cite{google-nature-2019}\cite{morvan2023phase}. By comparing the simulated results with those obtained from quantum devices, researchers can identify and mitigate potential errors or deviations, ensuring the reliability and accuracy of quantum computations. Moreover, classical simulators provide back-ends for developing quantum algorithms and demonstrating quantum advantage. These factors make classical simulation a crucial scientific computing task. On the other hand, basic techniques in classical simulators are mainly based on high-performance linear algebra, which is highly compatible with deep neural networks and many traditional scientific computing problems. This means that the development of classical simulation can naturally be generalized to many fields. As a result, the importance of efficient classical simulators has attracted much attention in the fields of quantum physics and computer science. 

Random quantum circuit (RQC) is one of the most attractive quantum hardware to achieve quantum advantage beyond classical computers. RQCs are artificially designed to generate highly entangled quantum states with exponentially increasing classical complexity, representing the most challenging tasks for classical simulation. RQCs have attracted much attention recently since  methods for simulating RQCs can provide reference or be directly transferred to classical simulations for other circuits. However, with the rapid development of RQCs in recent years, the time complexity of classical simulation is increasing at a rate of nearly $10^3$ times every two years\cite{morvan2023phase}. In comparison, peak performance of supercomputers approximately doubles every two years\cite{TOP500} since TITAN announced in 2011. The gap seems to announce the establishment of quantum advantage in these artificially designed tasks. 

Tensor network contraction (TNC) provides new possibilities for simulating large-scale circuits. Considering that traditional state vector (SV) simulators\cite{wille2019ibm}\cite{li2021sv} are strictly limited by memory demand, which grows exponentially with the number of qubits. A state vector of the 56-qubit zuchongzhi circuit\cite{zuchongzhi} requires 512 PB memory, which exceeds the capacity of any supercomputer. In comparison, tensor networks, which only need GB-level memory, have emerged as a promising approach to simulate large RQCs. 

TNC performs a classical simulation by representing a quantum circuit as a tensor network and sequentially contracts it to a single node. The powerful representation ability of tensor network allows it to efficiently capture and compress the entanglement structure of quantum states, which enables us not to store the whole quantum state throughout the computation. Slicing technique\cite{chen2018classical} endows it with memory adaption for different architecture and embarrassing parallelism. Moreover, since the backend of tensor contraction involves permutation and General Matrix Multiplication (GEMM), which is similar to neural networks, TNC aligns well with AI-hardware trends\cite{wang2019benchmarking}\cite{jouppi2017datacenter}. 

Although offering significant advantages, TNC faces several challenges that need to be addressed. To sum up, there are three main points: computational complexity, parallel overhead and floating point efficiency. As both depth and number of qubits are growing fast, recent RQCs\cite{morvan2023phase} has achieved even 6 orders of magnitude complexity increase. Finding better contraction paths\cite{gray2021hyper} can relieve this problem. Parallelism of TNC comes from slicing\cite{chen2018classical}, which brings overhead when offering exactly independent subtasks. For large circuits\cite{zuchongzhi2}\cite{morvan2023phase}, the overhead will rapidly expand to hundreds times over the original computation amount. \cite{huang2020classical} reveals the structure of \emph{stem} in TNC, where a large tensor sequentially absorbs small tensors. This is usually mapped into a series of narrow matrix multiplications leading to low FLOPS efficiency\cite{liu2021closing}\cite{pan2021simulating}. Fused design\cite{lifetime} is proposed to reduce memory access. But it has not completely changed the reality of memory access bottlenecks and still needs ultimate optimization. In the following sections, we will further show some new challenges in details when quantum circuits get complicated, which is an example of "quantitative change produces qualitative change". 

In this work, we proposed a series of strategies to further explore the performance improvement for TNC, mainly focus on parallel overhead and efficiency. On the process level, two data reuse schemes are designed to reduce the floating point operations. By carefully merging subtasks and organizing memory, we can eliminate slicing overhead to a low level with no harm to parallelism. On the thread level, a new step fusion based on core-array cooperation is designed. Different from core-independent fusion\cite{lifetime}, our new design treats the core-array as a whole and make full use of remote memory access (RMA) to further improve computation intensity by more than 2 times. As the computational overhead gradually becomes non-negligible, in-kernel vectorized permutation is proposed. Compared with the work \cite{li2021sw_qsim} that treats the permutation as a single kernel, our in-kernel circumstance meets more restrictions without stride-DMA support. We proposed more fine-grained design for different cases and reduced the number of vectorized operations to be much lower than \cite{li2021sw_qsim}. For new shapes of contractions in large circuits, we designed split-K operator with careful architecture mapping. Our efforts can accelerate the simulation of Zuchongzhi-60-24 circuit (60 qubits, 24 cycles)\cite{zuchongzhi2} by more than 10 times, and scale to more than 1024 Sunway nodes (399,360 cores).

Major contributions of this work include:
\begin{itemize}
\item Two data reuse strategies to reduce slicing overhead by 5 times on the process level.
\item An extended fused operator by core-array cooperation to improve data locality, resulting in further decrease in memory access.
\item A series of fine-grained optimizations and new operators to deal with new challenges for large circuits and computational-intensive status.
\item The simulation of Zuchongzhi-60-24 circuit is accelerated by more than 10 times, using up to more than 1024 Sunway nodes (399,360 cores).

\end{itemize} 

\section{Background and Related Work}

\subsection{Challenges of Large Tensor Networks}
Tensor networks (TNs) are often depicted as undirected graphs, where nodes represent tensors and edges connect tensors that share common indices. The contraction operation between two tensors is defined by the Einstein summation convention. In practical implementations, contraction is organized as a matrix multiplication operation of dimensions $M\times K \times N$, where the common indices are permuted to align along the dimension $K$. In the graphical representation of tensor networks, the contraction operation corresponds to the elimination of the shared edge between the two tensors, resulting in a new tensor with the remaining uncontracted indices. In a tensor network contraction (TNC) process, quantum circuits are represented as tensor networks, and tensors will be contracted along each edge until there is only one left.

Large tensor networks, transformed from large circuits, bring new challenges, including but not limited to extreme complexity, significant slicing overhead, new computation kernel and new hotspot distribution. 

\begin{figure}[htbp]
\centerline{\includegraphics[width=0.42\textwidth]{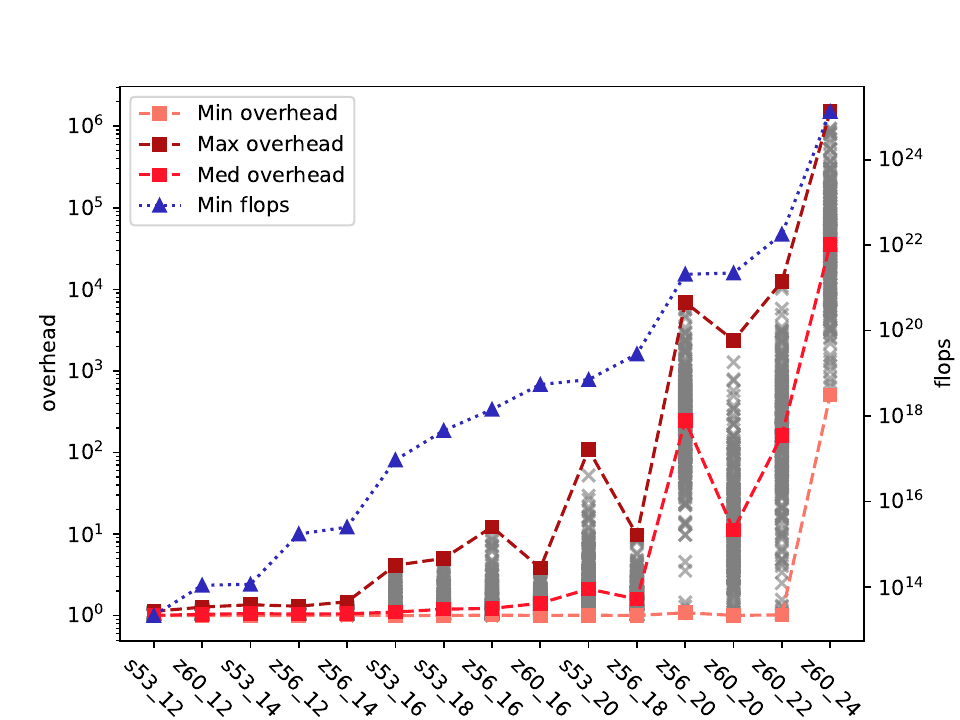}}
\caption{\textbf{Distribution of slicing overhead of different circuits.  Sycamore-53\cite{google-nature-2019}, Zuchongzhi-56\cite{zuchongzhi} and Zuchongzhi-60\cite{zuchongzhi2} are chosen for testing. Circuits are represented as name + qubits + cycles (s53\_12 denotes to Sycamore-53-12), and sorted by its minimum FLOPs. Memory limitation is set as rank-31. For each circuit, we searched 300 paths.}}
\label{distribution}
\end{figure}

Slicing is a key technique to establish the advantages of TNC over other methods whose memory demand grows exponentially with qubits. Dozens of indices are removed to reduce the maximum size of the intermediate tensors during TNC process, accompanied by additional overhead. Slicing overhead is concerned early in previous works\cite{gray2021hyper}. \cite{lifetime} proposed a concept, \emph{lifetime} to detect how a sliced index affect. Previously, slicing overhead did not attract much attention, as it was small enough to be ignored. However, things have changed for larger circuits, as the Fig~\ref{distribution} shows. Though for circuits like Zuchongzhi-56-20\cite{zuchongzhi} and Zuchongzhi-60-22\cite{zuchongzhi2}, a low overhead less than 10 can still be obtained, to search for these paths need long time, and these paths with low slicing overhead may not have a low total FLOPs. Zuchongzhi-60-24 is a critical point. Even the lowest overhead is up to $100+$. High overhead in one hand comes from the inherit complexity of large circuits, and the impact of the new structure is also prominent.

\begin{figure*}[htbp]
\centerline{\includegraphics[width=1.0\textwidth]{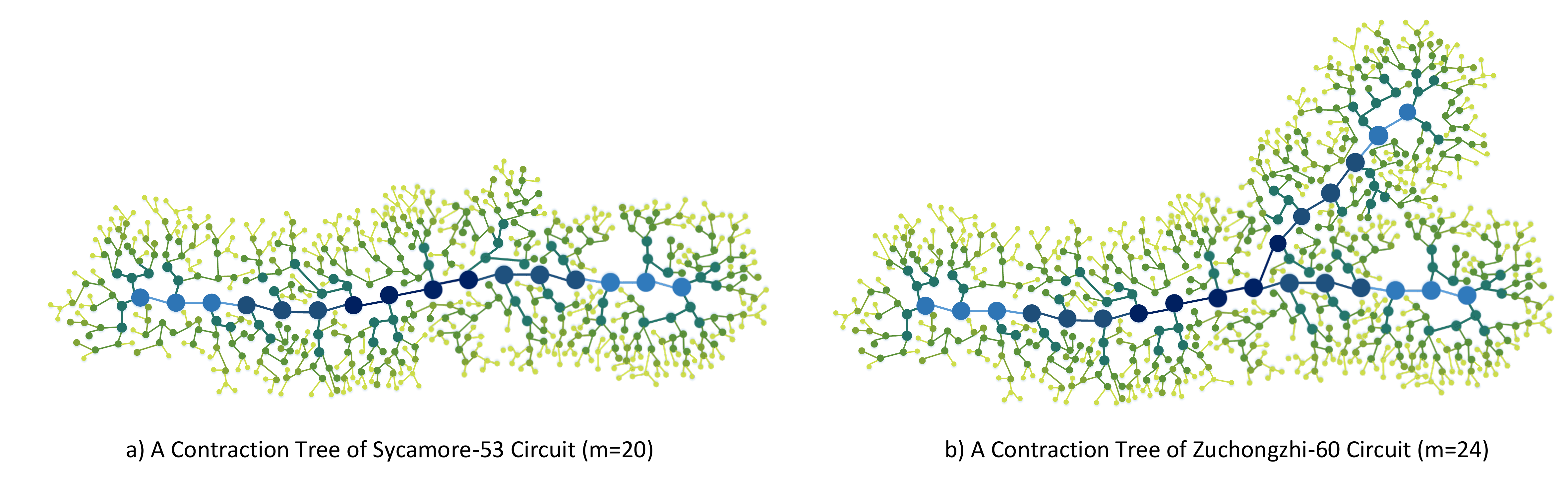}}
\caption{\textbf{Visualized structure of contraction trees. Each node represents a contraction step, and darker nodes indicates higher complexity. 300 contraction trees are searched for each circuit and one typical case is shown. a) Sycamore-53-20 circuit\cite{google-nature-2019}. b) Zuchongzhi-60-24 circuit\cite{zuchongzhi2}.}}
\label{multi-stem}
\end{figure*}

Larger circuits shows different structures on their contraction trees. Stem structure, as Fig~\ref{multi-stem} a) shows, is observed in \cite{huang2020classical} on Sycamore-53\cite{google-nature-2019} contraction tree, and extended in \cite{lifetime}. Here we follow the latter definition. As an almost linear structure, on the stem, a large tensor is contracted with small tensors sequentially. However, stem is not a general structure. There will be multiple stems detected for more complicated circuits, as Fig~\ref{multi-stem} b) shows. Multi-stem does harm to slicing. According to our experiments, most of contraction trees of Zuchongzhi-56-20 and Zuchongzhi-60-24 have multi-stem structure, while stem dominates in Sycamore circuits and low-depth Zuchongzhi circuits. Recall Fig~\ref{distribution}, we can see, multi-stem are correlated to high slicing overhead. 

Moreover, multi-stem requires us to support different contraction kernels. If two large tensors are contracted to a large tensor, that will be a near square-like GEMM, which is well-supported\cite{li2021sw_qsim}. However, if the result is a small tensor, the existing SWTT libraries\cite{liu2021closing} cannot handle efficiently. This leads to $K \gg M, N$ in a $M \times K \times N$ GEMM, which does harm to 2D-partition parallelism, and suffers from extreme memory bound. 

Inside the computational kernels, new hotspot distribution also brings chance. Fused design\cite{lifetime} significantly improved computation intensity by fusing several contraction steps together to reduce memory access. As a result, the hotspot distribution of TNC partially shifts from memory access to computation. In the step-by-step paradigm, computation is almost overlapped, which makes no sense for optimizations. As the tide of memory access recedes, computation and permutation dominate the process time in some cases. The new status quo requires us to improve both memory access and computation, instead of only optimizing the former one.

\subsection{Selected Architecture}
As a heavy time- and memory-consuming part, performing the actual contraction generally requires support from sophisticated supercomputers. A Sunway supercomputer based on SW26010Pro Processor is selected for this work. 

The major computing capability of a Sunway SW26010Pro chip is provided by 6 core groups (CGs), on each of which a managing processing element (MPE) and an 8 by 8 grid of computing processing elements (CPEs) are arranged. Each CG contains 16GB of main memory, and each CPE has 256KB of local data memory (LDM). To manipulate large intermediate tensors, an all-chip shared mode can be used to unite the main memory of the 6 CGs together. Direct memory access (DMA) with a bandwidth of 51.2 GB/s per CG is provided between the LDM and main memory. Due to the enormous arithmetic intensity, the memory access bottleneck often becomes the critical problem for optimization.

Remote memory access (RMA) with a peak bandwidth of more than 800 GB/s is designed for data exchange between CPEs within one CG. Unlike MPI\cite{walker1996mpi} and register communication on Sunway TaihuLight\cite{fu2016sunway}, RMA-based communication does not require a send-receive pair. Instead, it provides interfaces similar to DMA, which means we can adopt a more asynchronous design. However, the high bandwidth advantage of register communication is lost at low granularity, which prevents us from engaging in fragmented communication.

\subsection{Related Works}
The difficulty level of RQC simulation is one of the key indicators for the development of quantum computing. The first shot of "Quantum Supremacy" declaration was fired by Sycamore-53 circuits\cite{google-nature-2019} in 2019. In the following years, Zuchongzhi-56\cite{zuchongzhi}, Zuchongzhi-60\cite{zuchongzhi2} and Sycamore-70\cite{morvan2023phase} circuits were successively proposed. These circuits stand for the most complicated RQCs, bringing together all the difficulties that can be encountered when simulating RQCs. In \cite{morvan2023phase}, the number of qubits grows to 70, corresponding to an 8ZB size state vector. This size is far beyond the memory, even the hard disk of every supercomputer system. State vector and density matrix based simulators are restricted by exponential memory demand. TNC exists to be a potential method for these circuits with a large number of qubits.

Various TNC-based simulator frameworks are proposed, promoting the progress of simulation in different aspects. QFlex\cite{villalonga2020establishing} provided a well-built framework for the TNC process, while it is not that efficient for non-square contractions. Quimb\cite{gray2018quimb} and Cotengra\cite{gray2021hyper} made contributions to the simplification of the tensor network and the search for contraction paths. Alibaba developed slicing\cite{chen2018classical} to customize the memory demand with architecture, with additional computation. Some following efforts\cite{huang2020classical}\cite{lifetime} reduced slicing overhead to 1.x times the original complexity. AC-QDP\cite{zhang2019alibaba} and its following work\cite{huang2020classical} detected the 'stem' structure, which implies that narrow GEMM with extreme $M$ or $N$ dominates the TNC process. On GPU, Pytorch\cite{paszke2019pytorch} is developer-friendly but lack of fine-grained optimization for some edge cases. CuQuantum\cite{cuquantum} exists as a interlayer with composable primitives, supporting easy-to-use interfaces and flexible integration with high performance backends. SWTT\cite{liu2021closing} integrated techniques of previous works on Sunway architecture, and exactly calculated an amplitude in seconds. Lifetime-based methods\cite{lifetime} further pushed the limit by fusing several contraction steps together, and changes the computing paradigm of step-by-step contraction. However, this fused method to date only exists on Sunway architecture.

Efficient TNC simulators are also inseparable from the support of high performance tensor contraction libraries. As mentioned above, tensor contraction consists of permutation and GEMM, which are both classic subjects of high performance computing. Thus, general linear algebra libraries, like BLAS, LAPACK\cite{anderson1999lapack}, cuTensor\cite{cutensor} can be directly applied as backend. In particular for GEMM, cuTLASS\cite{CUTLASS} and SWQsim\cite{li2021sw_qsim} stand for state-of-the-art implementation on GPU and Sunway, respectively. If take permutation into account, optimizations for tensor contraction and GEMM still remain gap. Noting that memory access accounts for the nonnegligible time cost, Transpose-Transpose-GEMM-Transpose (TTGT)\cite{springer2018design} tried to fuse permutation and GEMM in one kernel. Tensor shape has a decisive influence on GEMM efficiency, and this influence is also inherited by TTGT. In \cite{goto2008anatomy}, GEMM is divided into 8 cases based on the relative sizes of $M$,$N$ and $K$, where $large \times large$, $large \times small$, $small \times small$ are named matrix (M), panel (P) and block (B), respectively. On Sunway architecture, \cite{liu2021closing}\cite{li2021sw_qsim} implemented efficient GEMM, GEBP, GEPB, and GEPP. In high-qubit-count circuits with multi-stem, GEPDOT often exists and accounts for more than 40\% time cost due to extremely low efficiency. \cite{li2021sw_qsim} realized vectorized permutation by single instruction multiple data (SIMD). However, in that work, instead of being a part of TTGT, permutation is organized as an individual kernel, and they only considered some simple cases. We should extend vectorized permutation inside the TTGT and fused kernel, with more general support.

Trade-off between time and space is a typical topic in computer science, but data reuse is still new to TNC. \cite{pan2021simulating}\cite{kalachev2021multi} has discussed data reuse between multi-amplitudes. \cite{liu2024verifying} further applied \emph{lifetime} to detect the chance of reuse. They identified the repeated computation and stored some intermediate results in free memory. Exactly the same computation occurs between not only amplitudes, but also slicing subtasks. \cite{lifetime} pointed out that slicing overhead comes from redundant calculation, and tried to alleviate it by finding better slicing indices. However, keeping slicing indices unchanged and tying to reuse data may be a more direct solution.

\section{Data Reuse Strategy}
\subsection{Overview of the Reuse Scheme}
In a TNC process, slicing strategy is applied to reduce memory demand, but meanwhile introduces overhead. According to \cite{lifetime}, the contractions uncovered by lifetime of sliced indices will be redundantly calculated. This provides a qualitative interpretation. To see how each index contributes to the final overhead quantitatively, here we define \emph{index overhead} as:

\begin{equation}
    O(a) = \frac{2C_{a}}{C_{ori}}
\end{equation}
In the equation, $C_{ori}$ is the complexity to contract a certain tensor network. After slicing index $a$, there will be two subtasks, each of which represents a component of $a$. $C_{a}$ is the complexity of one subtask. Thus, if $O(a) > 1$, i.e. $2C_{a} > C_{ori}$, there will be overhead after slicing $a$. Index overhead will be a key indicator when choosing slicing indices.

Slicing-generated subtasks provide extremely high parallelism at process level. Considering Sycamore-53-20 circuit\cite{google-nature-2019}, at least 22 indices will be sliced if the maximum rank is set to 31. Then there will be $2^{22}$ independent subtasks. For Zuchongzhi-60-24\cite{zuchongzhi2}, the number will be further $\sim 2^{45}$. Though the peak performance of supercomputer has grown to Exa-scale, the number of processes did not follow the same trend. According to \cite{TOP500}, the leading systems, like Frontier(rank-1), Fugaku(rank-1 at 2021), Summit(rank-1 at 2018) and Sunway TaihuLight(rank-1 at 2016) have accelerators/nodes at the order of $10^5 - 10^6$ for process level parallel (estimated by the ratio between peak performance and per-device/per-node performance). That means that there will be multiple subtasks executed sequentially on each process. By proper organization, these subtasks can be encoded by some sliced indices with high index overhead. Data reuse happens between these inner-node indices, without interprocess data exchange.

Essentially, data reuse trade computation as storage, which requires careful memory manipulation. After slicing, a maximum tensor-rank $m$ can be maintained not to exceed memory capacity. For RQCs, with all indices of size 2, the memory cost for a contraction $A \times B \rightarrow C$ is at most $3*2^{m}$, where $m$ is the highest rank, and the cost varies by a factor of 2 with different values of $m$. This property may lead to memory waste for some architectures since the gap will be too large for big $m$. However, this discrete memory distribution often does not align well with the available capacity. Global memory of an Nvidia A100 GPU is $40GB$ or $80GB$, where $40\%$ of the memory is not utilized. Similarly on Sunway 26010Pro, $50\%$ of the $96GB$ main memory may remain idle.

These idle memory resources presented an opportunity for data reuse strategies, which can reduce slicing overhead for large tensor networks by trading off memory for computation. By merging some computations and reusing intermediate results, a significant amount of overhead can be eliminated.


\subsection{Tree-like Reuse}

Chance of data reuse comes from slicing-caused repeated calculation. Lifetime\cite{lifetime} provides analysis of the source of overhead. Before lifetime starts, the sliced subtasks will do exactly same work. After lifetime ends, computation in all subtasks can be described as $\sum_i A_iB$ which leads to massive complexity compared with $(\sum_i A_i)B$ by distributive law of Einstein summation. Our target is to store and reuse the intermediate result from other subtasks and merge subtasks properly to utilize the distributive law. 

\begin{figure}[htbp]
\centerline{\includegraphics[width=0.46\textwidth]{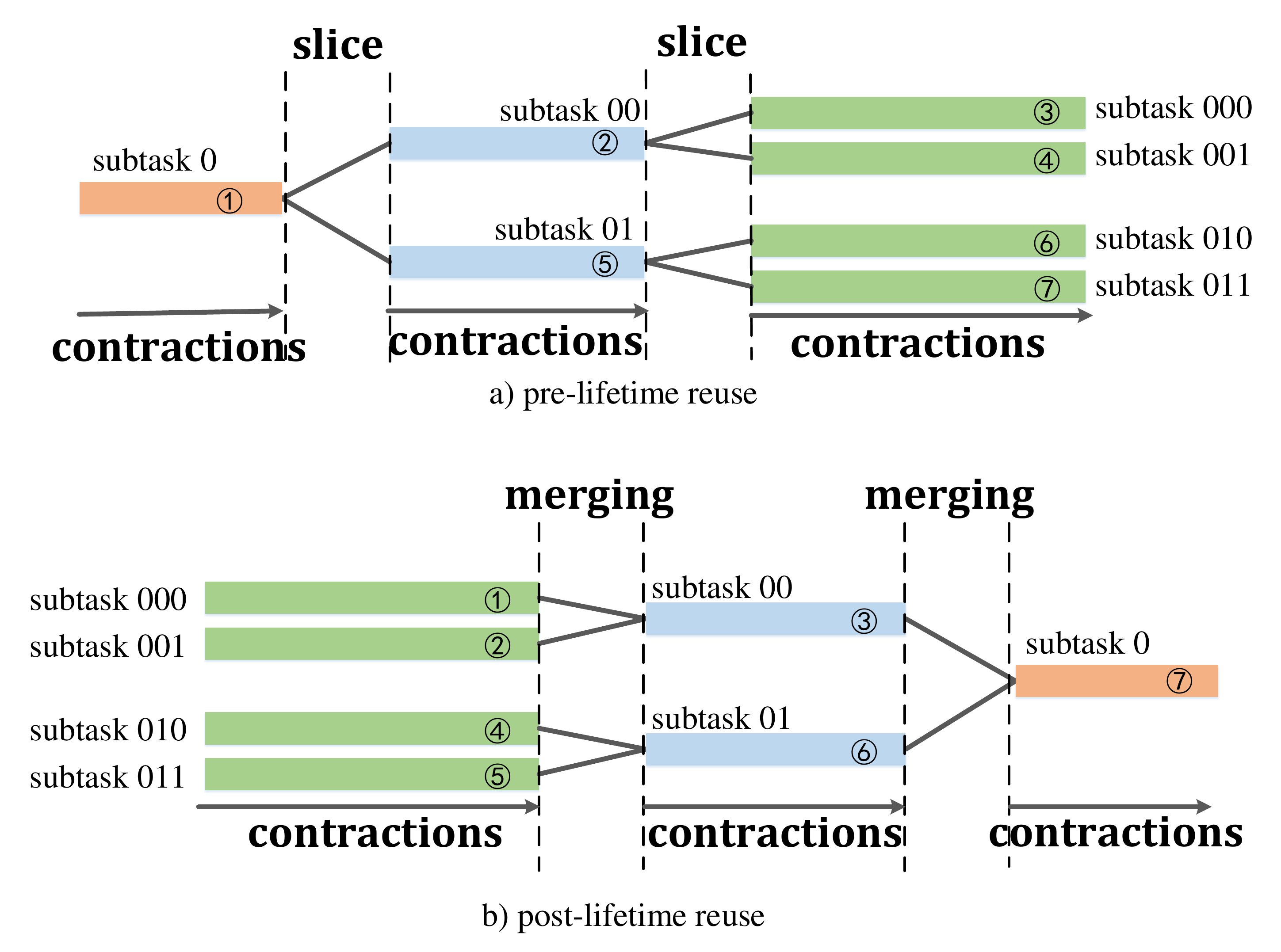}}
\caption{\textbf{Tree-like data reuse. In both sub-figures, the sequence number represents the order of execution, and the thick solid lines denotes to continuous contraction steps. In a), slicing happens when lifetime of an index starts, and every subtask will be split in two. Intermediate tensors will be stored when a thick line ends, and deleted when both of its two children start. In b), merging works at the end of lifetime, and the two subtasks alongside the corresponding dimension will be reduced. When both of two subtasks paired on a index ends, the intermediate tensor stored by the first one can be deleted.}}
\label{tree-reuse}
\end{figure}

We first consider pre-lifetime reuse. For two subtasks with different only on the projection of one sliced index $a$, their computations before lifetime $a$ starts are exactly same. The last same intermediate tensor can be stored in the memory when calculating the first subtask. When the first subtask is finished, the second subtask can simply start from the stored intermediate tensor. Generalizing to the condition of $n$ sliced indices, the reuse scheme will form a binary tree, as Fig~\ref{tree-reuse} a) shows. To achieve the minimum storage demand, a depth-first traverse is applied. Following the traverse order, one should only store one tensor at each fork, i.e. the number of stored tensors is equal to $s$, when $s$ is the sliced indices within the process. 

For post-lifetime, there will be a similar reuse scheme. If the lifetime of $a$ ends, the first subtask can simply stop and leave the intermediate tensor in memory. When the second subtask get to the same position, we sum up the result tensors to merge these two subtasks and continue the following contractions. The depth-first strategy also works here, the order is shown in Fig~\ref{tree-reuse} b). The storage situation is also the same as above. 

Though not all intermediate tensors meet the maximum size, storing too many tensors may exceed the memory. Here we proposed a dynamic tuning method to customize the memory demand for architecture. According to \cite{liu2024verifying}, the rank of intermediate tensors on the stem are distributed as a hill, with a process of rising first and then falling. If the stored tensors are too large for the idle memory, we can do earlier slicing and later merging to store smaller tensors, at the price of slicing overhead. This overhead is much smaller than the original one.

\subsection{Spindle-like Reuse}

\begin{figure}[htbp]
\centerline{\includegraphics[width=0.50\textwidth]{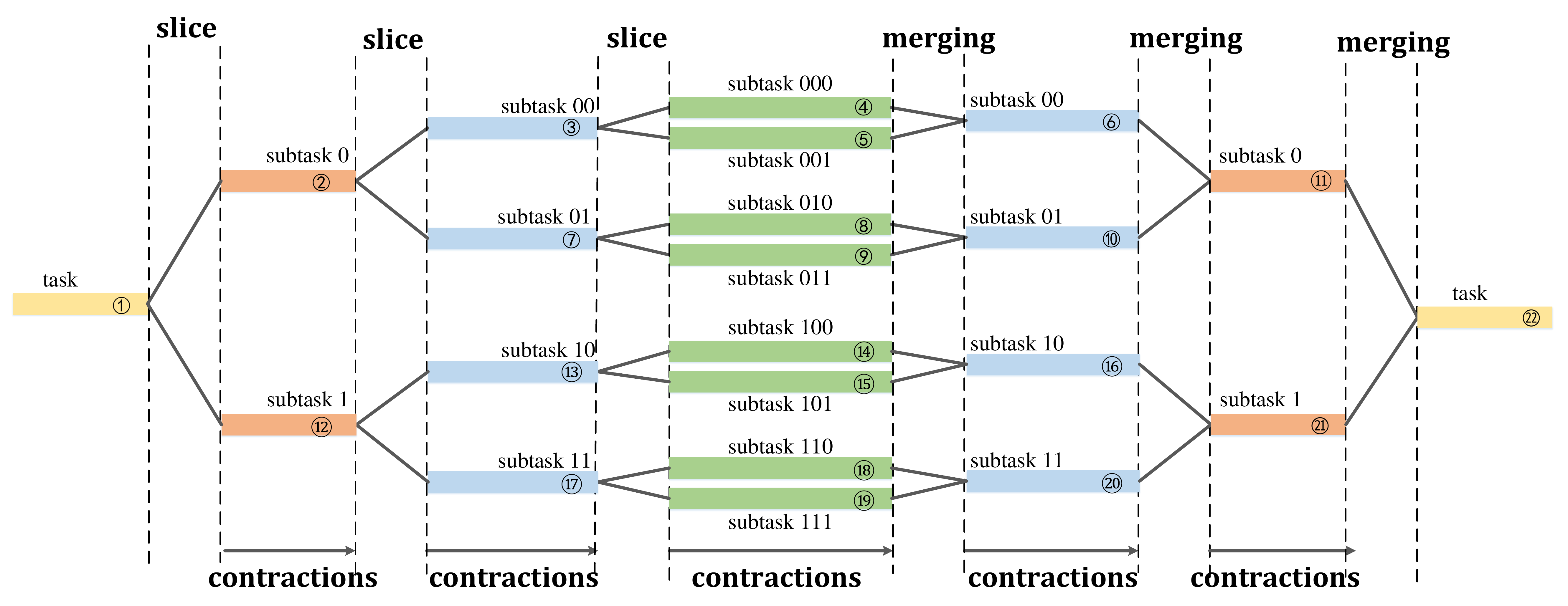}}
\caption{\textbf{Spindle-like reuse, which is the combination of pre-lifetime and post-lifetime reuse. The sequence number represents the order of execution, and the thick solid lines denotes to continuous contraction steps. Intermediate tensors are stored and deleted following the rules of both pre-lifetime and post-lifetime reuse.}}
\label{spindle-reuse}
\end{figure}

Tree-like reuse can largely reduce slicing overhead. To fully utilize data reuse, pre-lifetime reuse and post-lifetime reuse should be combined as a spindle.  As Fig~\ref{spindle-reuse} shows, inside a process, all subtasks are organized as a spindle, with all the repeated calculation removed. Every horizontal line denotes to a subtask with a series of contractions. Binary fission and merging happens when lifetime of a sliced index start or end. Naturally there will be memory issue since too many intermediate result should be kept. To store as fewer tensors as possible, we designed a two-way depth-first traverse strategy. Serial numbers in Fig~\ref{spindle-reuse} denote to the execution order. We first go depth-first to a leaf node (subtask 000) of the pre-lifetime reuse tree with 3 tensors stored. After subtask 000 is finished, the result is stored for merging, and subtask 001 starts. At this time, the tensor generated by subtask 00 can be deleted since it will not be used any more. Moreover, the tensor stored by subtask 000 will also be removed after merging, with the result of merged subtask 00 stored. As we can see, under our two-way traverse, the combination of two reuse strategies only brings one additional tensor (generated by leaf subtasks for merging), which is very friendly for storage.

However, spindle-like reuse may be invalid if sliced indices show an irregular distribution, since it requires lifetime of sliced indices are nested as Fig~\ref{spindle-reuse} shows. This condition is not that common in practice. Here we applied the branch exchange introduced in \cite{huang2020classical} to solve this problem. When a large tensor $A$ is doing continuous contractions with small tensors $B_1$ and $B_2$, swaping the contraction order of $B_1$ and $B_2$ will not affect the complexity. Based on this observation, we can tune the lifetime of sliced indices for lower overhead and a nested structure.

\section{Fusion with Core-Array Cooperation}
Fused design\cite{lifetime} is proposed to reduce memory access in the TNC process, especially for stem\cite{huang2020classical} where a large tensor sequentially absorb small tensors. A fused section of length n can cancel $2(n-1)$ times of intermediate memory access (load and store), with only the first load and the last store remaining. Length is a significant parameters of fused design, which directly determines the amount of memory access. Previous methods greatly increased the computational density, but did not guarantee that all cases will be converted into computationally intensive problems. In this work, we will achieve this by lengthening the fused section.

In the previous design, length of fused section is strictly limited by memory capacity. 256-KB LDM on Sunway architecture can only handle rank-13 tensors, and the fused section have to terminate if the index needs to be contracted is not organized inside LDM. The missing indices should be reloaded into the LDM through additional memory access operations to restart a new fused section. If we can expand the ranks that each CPE can handle, reorganizing fused section will be less frequent, leading to less memory access.

\subsection{Communication Scheme}
\begin{figure}
    \centering
    \includegraphics[scale=0.27]{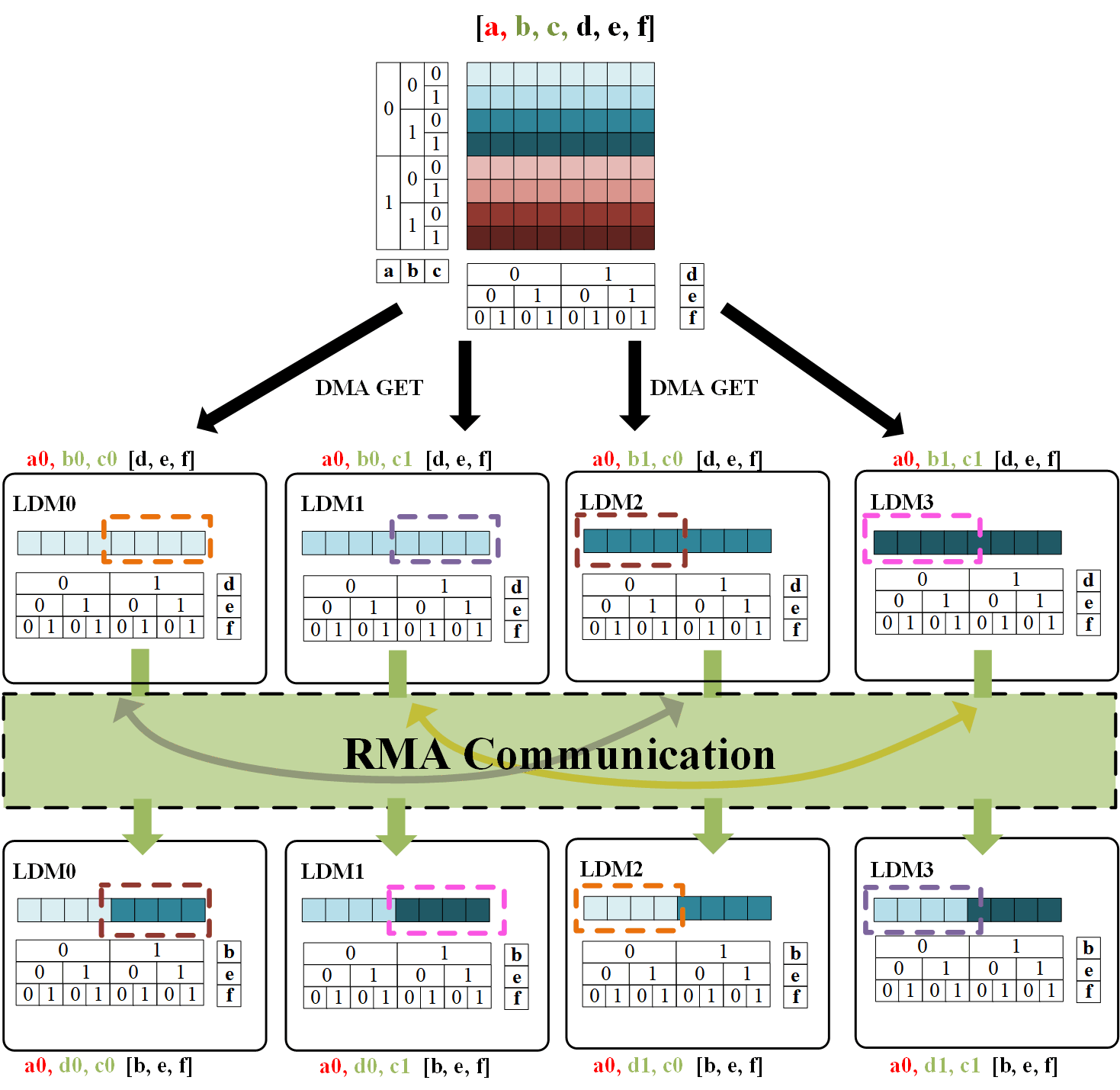}
    \caption{\textbf{RMA communication scheme to lengthen the fused section. Without loss of generality, here shows 4 CPEs cooperation. A rank-5 tensor is distributed in 4 CPEs, with 3 indices intra-LDM and 2 indices inter-LDM. When the inter-LDM indices need to be contracted, RMA works to swap indices. After communication, an inter-LDM index and an intra-LDM index exchanged positions.}}
    \label{communication_scheme}
\end{figure}

Core-Array Cooperation helps break through rank-13 memory limitation. With RMA communication, the local memory of 64 CPEs can be organized to store rank-19 tensors. The additional indices are hidden between CPEs, called inter-LDM indices, as Fig~\ref{communication_scheme} shows. Inside LDM, there is still a rank-13 tensor, whose indices are named intra-LDM indices. As a result, indices of the original large tensor is divided into 3 parts: sliced indices, inter-LDM indices and intra-LDM indices.

The communication scheme is described in Fig~\ref{communication_scheme}. Before an inter-LDM index $b$ is about to be contracted, we should convert it to a intra-LDM index. The 0 components and 1 components of $b$ is distributed in different two-CPE pairs, respectively. In Fig~\ref{communication_scheme}, CPE 0 and CPE 2, CPE 1 and CPE 3 each form a pair (components of indices except $b$ should be same, as CPE 0 and CPE 2 both hold the 0 component of $c$). Then, an intra-LDM index is chosen for exchange. Here we simply use $d$ for largest granularity. In total, two half-tensors will be transported in the point to point communication. After communication, $b$ is converted to an intra-LDM index, while $d$ becomes an inter-LDM index.

Choosing the best inter-LDM indices helps to reduce the amount of communication. At the beginning of fused sections, we will reorganize the tensor to put the indices with the longest lifetime between CPEs. During communication, the index with the longest remaining life, i.e. lifetime starting from current step, will be selected for exchange. These strategies ensures that communication will not happen at most contraction steps.

\subsection{Data Transaction Merging}
Though we have chosen the indices with longest lifetime as inter-LDM indices, we need to do at most 6 time RMA communication in one kernel, since each communication can only swap two indices. Actually, we can enlarge the communication group to swap more indices with lower cost.

When lifetime of two inter-LDM indices end at the same contraction step, we can organize a 4-CPE communication group instead of pairs. Two intra-LDM indices will turn to be inter-LDM after communication. For each CPE, $\frac{1}{4}$ of the correct data is already held by itself, and it should take $\frac{1}{4}$ tensor from each of the other 3 CPEs in the same group. Compared with the one-by-one swap strategy, this batch-swap reduces the average amount of communication per-CPE from one tensor to $\frac{3}{4}$ tensor.

This strategy can be extend to more indices. The size of communication group is exponentially related with $n$, the number of swapped indices, while the communication amount reduction ratio follows:
\begin{equation}
    \frac{Comm_{one\_by\_one}}{Comm_{batch}} = \frac{n/2}{1-\frac{1}{2^n}} = \frac{n2^{n-1}}{2^n - 1}
\end{equation}

So, if communication has to be done to swap one or two inter-LDM indices, we can swap more indices whose lifetime will end in the next several steps for less communication.

\section{Further Optimized Contraction}
With core-array cooperation, memory access is further reduced. In the new computation paradigm, permutation and GEMM dominates the process time. Considering that cGEMM kernel is already carefully optimized, permutation turns to be the next target. Besides in-kernel optimization, contractions with many common indices and few free indices are extremely inefficient in the present implementation, which needs specific design.

\subsection{Vectorized Permutation}
The purpose of permutation is to rearrange common indices to $K$ direction of the matrix. Then Einstein summation can be done by GEMM with high performance. That means we do not need to deal with all kinds of permutation. Instead, we should only consider one pattern where a select few indexes are relocated to the front or the back, while the remaining indexes maintain their original order. This pivotal pattern guarantees the existence of a long sequence of consecutive indexes during the permutation process. For convenience, we applied exchange to ensure all large tensors staying at the position as multiplier, where data locality will be kept as much as possible. 

\begin{figure}
    \centering
    \includegraphics[scale=0.72]{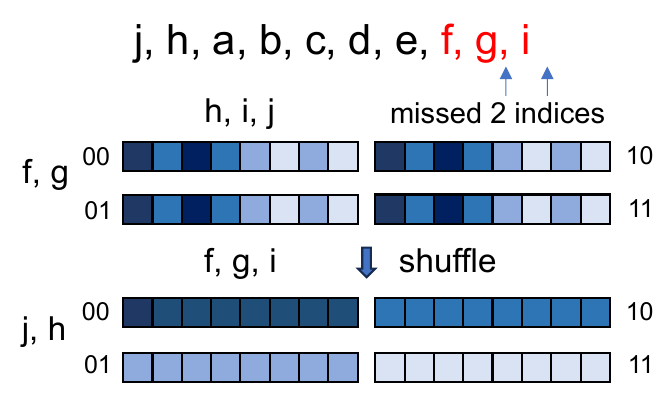}
    \caption{\textbf{An example of vectorized permutation. Red indices in the permutation map are continuous in the memory after permutation. Four registers involve shuffling. The registers load data continuously from the original tensor, and then shuffle to hold the three red indices.}}
    \label{permute}
\end{figure}

Here we proposed two parameters, stride and offset, for a certain permutation. The last index of a tensor directly effect on data continuity. If the last $m-1$ indices are rolocated to the front, offset is set as $m$. After permutation, the last index of the new tensor should be the $n-m+1$th index of the original tensor. In other words, $m=1$ means at least two complex number can be treated as a whole during permutation. Stride denotes to the length of the last and longest continuous subsequence in the index sequence after permutation. Intuitively stride represents the length of continuous parts.  Tensors in the stem has similar shape, that means the costs of permutations are close at each steps in the fused kernel. According to Amdahl's law, single instruction multiple data (SIMD) should be applied to all these permutations to avoid the drag of unparalleled parts.

\begin{figure}[htbp]
\centerline{\includegraphics[width=0.43\textwidth,height=0.4\textwidth]{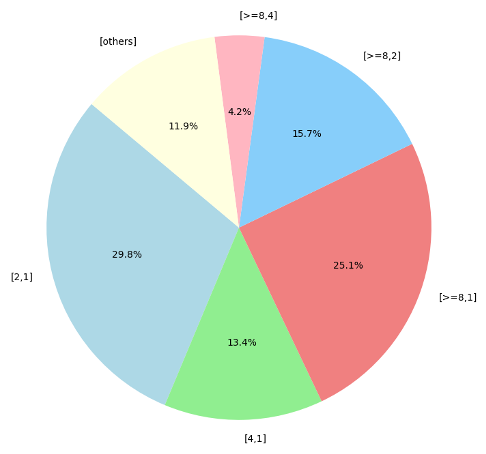}} 
\caption{\textbf{Proportion of various Permutation Situations, represented by [\(2^{strides}\), \(offset\)], in total permute time in Sycamore 53\_20}}
\label{vector}
\end{figure}

For single precision complex numbers, we use 512-bit vectorized registers to process 8 numbers in parallel. The ideal condition is when $offset = 1$ and $stride >= 3$, where permutation can be done only by load and store instructions. If one or more of the last three indices are relocated, we need cooperation from multiple vectorized registers. Fig~\ref{permute} illustrates a vectorized permutation by shuffle, with $offset = 2$ and $stride = 1$. In this circumstance, 4 registers will work together to hold all five indices (f, g, h, i, j).

For Sycamore-53 circuit (m=20)\cite{google-nature-2019}, the distribution of offset and stride are shown by Fig~\ref{vector}. Without shuffle, SIMD instructions can deal with only a quarter of circumstances. In most instances, there are 1-2 missed indices, where we do not need too much registers. For each CPE on Sunway architecture, there are only 32 vectorized registers. If most of them are needed to load data, we need to carefully allocate resources to accomplish shuffling.

Our vectorization strategy maintains a universal design while taking into account some special optimizations. The number of registers grows exponentially with the missed indices of the last 3 indices. Considering that a 4-index permutation can be done within one shuffle, with one index relocated, we can guarantee to finish permutation with at most triple shuffle (using 8 registers to load data). Configuration of shuffle instructions can be deduced by offsets and stride, instead of Manual design. 

\subsection{Split-Common TTGT}

SW\_BLAS and SW\_Qsim\cite{li2021sw_qsim} provides high performance square GEMM and GEPP on Sunway architecture. However, GEPDOT and the corresponding TTGT are not efficiently implemented. When simulating large circuits like Zuchongzhi-60 (m=24) \cite{zuchongzhi2}, though there are only several GEPDOTs among hundreds of contraction steps, GEPDOT accounts for more than 60\% of the total time. In order to prove the universality of this problem, we selected 200 contraction paths in each of Zuchongzhi circuit\cite{zuchongzhi}\cite{zuchongzhi2} with different structures and counted the number of GEPDOT occurrences.

\begin{figure}
    \centering
    \includegraphics[scale=0.22]{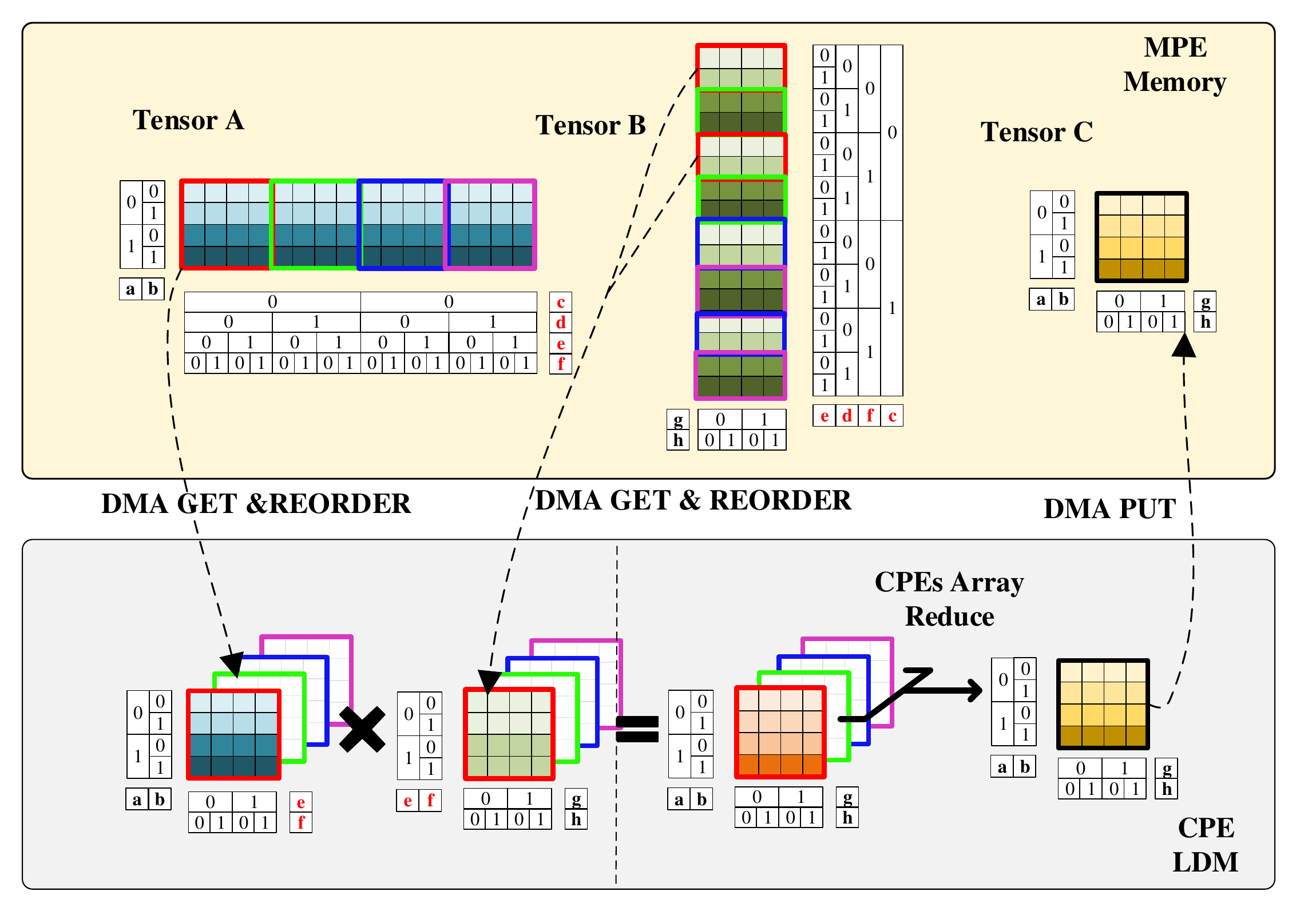}
    \caption{\textbf{ Split-Common TTGT tensor contraction. Fusion design of permutation with split-K matrix multiplication. All free indices are inside LDM in order to utilize high performance cgemm kernel. }}
    \label{splitk}
\end{figure}

Our split-common TTGT fused permutation and GEPDOT together, as Fig~\ref{splitk} shows. To full utilized the complex GEMM kernel, all free indices will be loaded into local memory by stride-DMA. Partition happens along the common indices. Permutation is vectorized by the strategies above. After computation, RMA-based reduction is applied to sum up the result. When the output tensor get larger, the problem is on the border between GEMP(GEPM) and GEPDOT, where full array reduction brings heavy communication and error probability. Under this circumstance, M(N) is not large enough for complete 1-D or 2D partitioning. As a result, we will do M(N)-K mixed partitioning. 1-D partitioning along M(N) direction divides the 64 CPEs into several communication groups, while the strategy in Fig~\ref{splitk} works inside each communication groups. This ensures that our method can smoothly transition to square-like GEMM.

According to \cite{higham2002accuracy}\cite{higham2022mixed}, floating point error of GEMM is:
\begin{equation}
    |\hat{C} - C| \leq \gamma_K |A||B|
\end{equation}
$\gamma$ is a scalar strongly related to K. To relieve this problem, we applied FMA instructions and blocked summation. As the output is small, we can allocate space for $b$ tensors, then do summation by blocks. With this strategy, $\gamma_K$ can be reduced to $\gamma_{\frac{K}{bg}}$, where $g$ is the size of communication group. 

\section{Error Mitigation and Correction}
Due to the huge computational space and time complexity of quantum circuit simulation based on tensor network, it is required to carry out large-scale parallel on Sunway supercomputer. However, with the grown of nodes, small problems of the network of communication, IO system, and the computation cores are more likely to be magnified. We mainly discuss two kinds of hardware error in this section: the problems which cause the program to crash, and the problems which influence the calculation results. In this work, we adopted a mechanism to support the retention and recovery function of variable parallel scale. 

The stability of large-scale communication challenges the reliability of the final result. To collect data from the whole supercomputer, we should stand a error rate as:
\begin{equation}
    ER_{all} = 1 - \prod_{i=1}^N (1 - ER_i) 
\end{equation}
where $N$ is the number of nodes, and will grow rapidly while $N$ increase even if $ER_i$ is a small amount. Considering that the calculation between the subtasks generated by slicing are independent except the only reduction at the end of the whole computation, we can apply IO or second level communication to reduce the scale of communication group. Considering about the fat-tree structure of the grid topology, in our design, $256$ nodes in a super-node are arranged as a communication group, and the collected result are write to files or stored at the $0-th$ node in the group, depending on the parameter provided by system operation and Maintenance. After all the calculation done, we collect up the results together. This design also reduces the destructive power of bad cores, and we will pay a smaller price when any kinds of error really occurs.

Moreover, to check whether there is wrong number in the results, we use a small-scale program which do the same computation with the large-scale one and only calculate hundreds of subtasks for verification. The small-scale program will be finished in seconds and provide parts of results of the large-scale one. We repeatedly run the small-scale one with different nodes and subtasks, and compare it with the results from the large-scale one.

\section{Evaluation}
The experiments are done on Sunway 26010Pro architecture. We use Zuchongzhi-56\cite{zuchongzhi} and Zuchongzhi-60\cite{zuchongzhi2} circuits (The most complex opensource circuits) with different cycles for test. Circuits are referred to as zcz\_n\_m, representing n qubits and m cycles. Contraction path is found by cotengra\cite{gray2021hyper} in 1 hr. All performance results are repeated by 10 times and average was taken. Except scaling results, other experiments are executed intra-node.

\subsection{Complexity Reduction}
\begin{figure}[htbp]
\centerline{\includegraphics[scale=0.5]{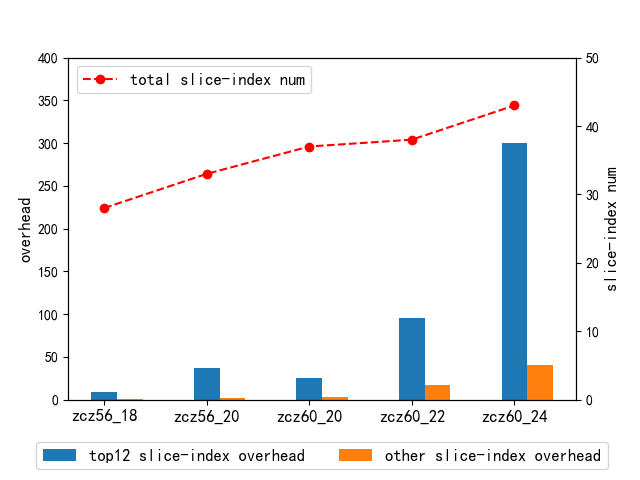}} 
\caption{\textbf{Overhead of top 12 sliced indices and the other indices. Overhead is calculated by the ratio between the sliced complexity and the original complexity. Contraction paths with low complexity are chosen for test.}}
\label{top12_index_overhead}
\end{figure}

Data reuse can significantly reduce time complexity. In our work, due to the limitation of memory, we can not make full use of data reuse. Instead, we can simulate the memory cost and the overhead of each index before actually executing the contraction. By this pre-process, indices with high overhead and acceptable memory cost are chosen for reuse. Fig~\ref{top12_index_overhead} shows the overhead of top 12 sliced indices and the other sliced indices. It indicates that the huge overhead is caused by a few indices, which verified our assumption that different index contributes differently during slicing. Furthermore, we can achieve a significant improvement by reusing several high-overhead indices, since only 12 sliced indices (4096 subtasks) accounts for major overhead. For the tested large circuits, the slicing set is often larger than 30, which means the remained indices can provide parallelism for more than 262144 processes (18 indices) with much lower overhead.

\begin{figure}[htbp]
\centerline{\includegraphics[scale=0.55]{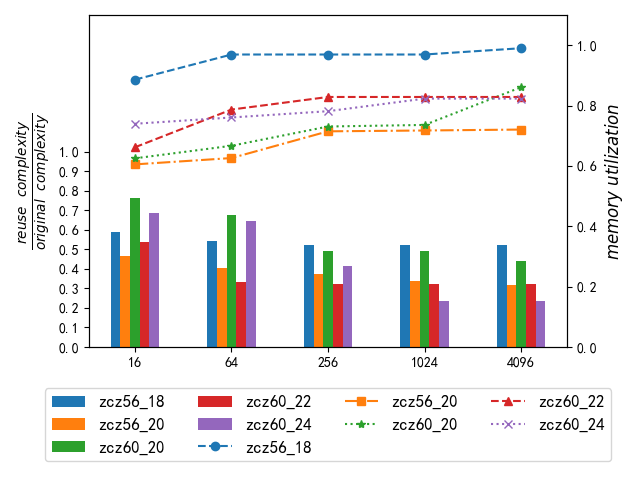}} 
\caption{\textbf{Result of data reuse. The histogram shows the overhead reduction for different circuits and different number of reuse subtasks. Number of reuse indices is just the logarithm of 2 for subtasks. The line chart shows memory utilization. For most circuits, there is still free space after reuse.}}
\label{reuse_overhead}
\end{figure}

Fig~\ref{reuse_overhead} illustrates the results of spindle-like data reuse. Reuse directly reduces the complexity and improves memory utilization. The general rule is that completed reuse will bring better result for more complex circuits, since slicing overhead is growing with circuit size. For some circuits, memory cost keep nearly unchanged mainly because there are several sliced indices end their lifetime at the same contractions, which leads to only one tensor stored. Another important reason is that, tensor-rank on a stem is also distributed like a spindle. At the two ends of a stem, where lifetime of most sliced indices start and end, the tensor is not that large. The closer we get to both ends, the tensor will be exponentially smaller. As a result, the memory cost will be slowly grow when we reuse more indices. In practical, this specific effect also depends on the contraction path. Since one more 16-GB tensor will bring great impact to the memory utilization, a carefully selected path may benefit a lot.

\subsection{Performance Results}
\subsubsection{Core-Array Cooperation Fusion}
\begin{figure}[htbp]
\centerline{\includegraphics[scale=0.5]{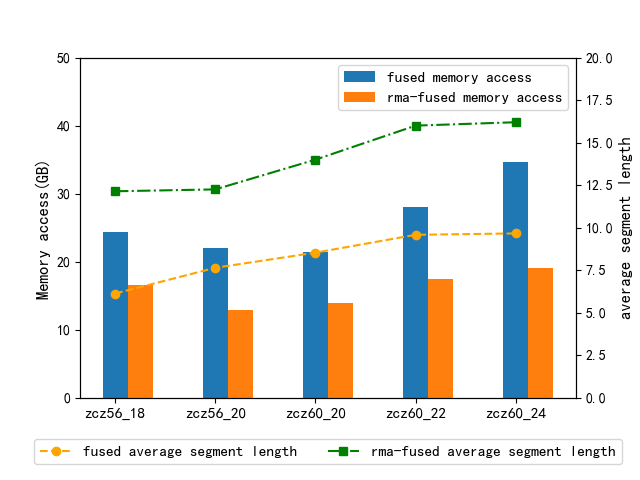}} 
\caption{\textbf{Amount of memory access and length of fused section. The average of multiple tests is shown.}}
\label{mem_access}
\end{figure}

With core-Array cooperation, memory access is replaced by faster RMA communication. Fig~\ref{mem_access} illustrates the effect of memory access reduction and the length of fused section. After optimization, both of memory access amount and fused length has gained a near 40\% improvement. The further reduction of memory access indicates that computational intensity of the fused kernel is further improved, and more attention should be paid on computation. Since our strategy does not affect the granularity, time cost of DMA will decrease proportionally.

\subsubsection{Permutation}
Table~\ref{vshuf} demonstrates the results of SIMD-optimized permutation. In our experiments, different configurations are set for different offset and stride. We tested the summed original time and optimized time for every case. For each test, we repeated 100 times and recorded an average result. Instructions include $vldd$, $vstd$ and $vshuf$, which are used for loading, storing data, and do shuffling. For {[2,1]}, there are two situations that need to be discussed in categories, with 9 and 10 instructions, respectively.

We carefully organized the instructions to reduce data dependence and enhance pipeline efficiency. Proper loop unrolling also helps a lot to cancel data dependence. Those are the most important reason why the speedup can breakthrough 8 times. There are still some cases which are hard to deal with or inefficient after vectorization. As a result, the speedup of total time is less than 8. Considering that the optimized permutation accounts for only 10\% of the whole TNC process, the marginal benefit of further fine-grained optimization is low. 

Compared with the previous work\cite{li2021sw_qsim}, our strategy can deal with more complicated conditions with much higher performance. Here we apply a performance model like what \cite{xu2018taming} did. Since permutation as a kernel is completely a memory bound problem, permutation itself can be fully overlapped by memory access, which leads to a full DMA bandwidth as 51.2GB/s per CG. Under this assumption, the permutation will be extremely inefficient with time even longer than memory access. If we make a loose assumption that no overlapping happens, the bandwidth of permutation itself will be less than 160GB/s for dispersed permutation, also much lower than the peak. In our work, the original approach can achieve a bandwidth of 514.46 GB/s, and the optimized one reaches 3526.82GB/s.

\begin{table}[h]
\caption{Improvement of permutation by vectorization.}
\begin{center}
\begin{tabular}{ccccc}
    \toprule
    \textbf{[\(2^{stride}\), offset]} & \textbf{ori time(s)} & \textbf{opt time(s)} & \textbf{instructions} & \textbf{speedup} \\ 
    \midrule
    {[2,1]} & 0.648 & 0.041 & 9-10 & \textbf{1597\%} \\ 
    {[4,1]} & 0.291 & 0.026 & 9 & \textbf{1093\%} \\ 
    {[4,2]} & 0.139 & 0.013 & 9 &\textbf{1043\%} \\ 
    {[$\geq$8,1]} & 0.546 & 0.059 & 8 & \textbf{918\%} \\ 
    {[$\geq$8,2]} & 0.341 & 0.047 & 9 & \textbf{728\%} \\
    {[$\geq$8,4]} & 0.091 & 0.013 & 10 & \textbf{722\%} \\
    Total time & 2.18 & 0.318 & NA & \textbf{683\%} \\ 
    \bottomrule
\end{tabular}
\end{center}
\label{vshuf}
\end{table}
\subsubsection{Split-common TTGT}
\begin{figure}[htbp]
\centerline{\includegraphics[scale=0.28]{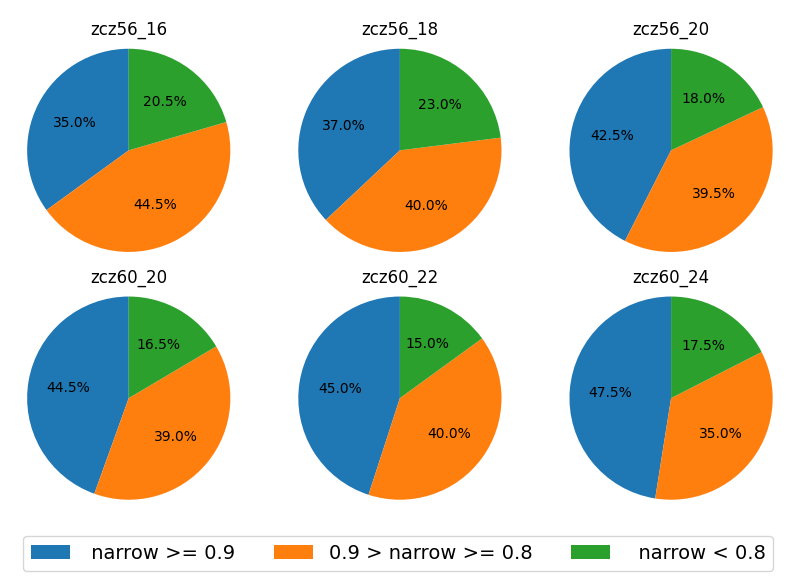}} 
\caption{\textbf{Narrow rate distribution for large circuits. For each contraction path, there are several GEPDOT steps. We counted 200 contraction paths to show the pattern.}}
\label{narrow_percentage}
\end{figure}

To describe the degree to which K dominates, we defined a factor $narrow = 2N_{common} / (N_A + N_B)$, where $N$ is the number of indices. Different from previous works in matrix multiplication, we use a definition based on indices, because extremely K-dominating GEPDOT exists with different number of common indices. Fig~\ref{narrow_percentage} demonstrates the distribution of GEPDOT in large circuits. As the circuits becoming more complex, GEPDOTs are growing to be more narrow. For the hardest circuits, near 50\% of GEPDOTs have a $narrow$ rate more than 0.9 (for a $30\times30$ contraction, there will be 27 common indices, and K will be $2^{23}$ times larger than M and N). That means it is necessary to design specific kernel for these extreme conditions.

\begin{figure}[htbp]
\centerline{\includegraphics[scale=0.5]{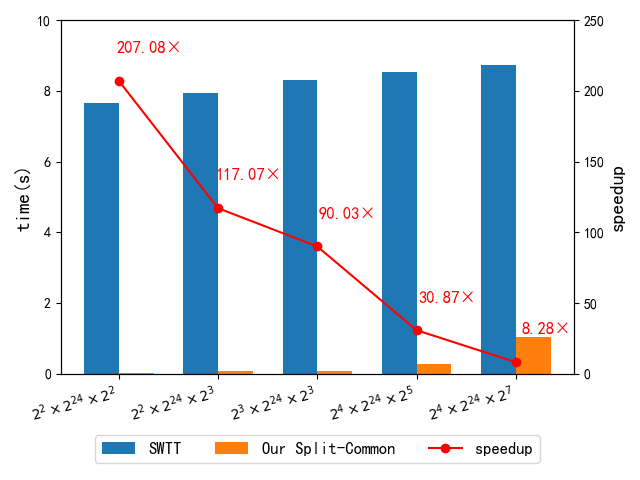}} 
\caption{\textbf{Performance results of split-common TTGT. Time costs are the average of multiple tests.}}
\label{splitk_performance}
\end{figure}

The performance result of our split-common TTGT is demonstrated as Fig~\ref{splitk_performance}. Here we fixed the number of common indices and tested cases with different free indices. Compared with SWTT library\cite{li2021sw_qsim}\cite{liu2021closing}, when $narrow \geq 0.9$, our split-common kernel can achieve more than 100 times speedup. With the reduction of $narrow$ rate, the speedup will decrease, but still maintain higher performance than SWTT.

\subsubsection{Scaling}
\begin{figure}[htbp]
\centerline{\includegraphics[scale=0.5]{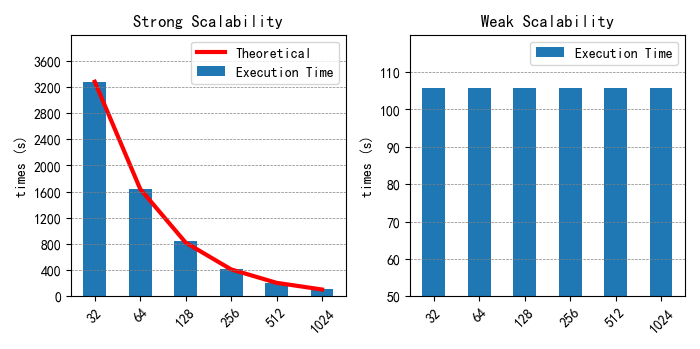}} 
\caption{\textbf{a)Strong scaling results. Number of subtasks is fixed as 16384. b) Weak scaling results. Number of subtasks increases with processes. All experiments are done on Zuchongzhi-60-24 circuit.}}
\label{scaling}
\end{figure}
We tested weak and strong scaling on the hardest circuit as Fig~\ref{scaling} shows. Since both of our process level and thread level optimization did not do harm to parallelism, scaling results performs similar with \cite{liu2021closing} and \cite{lifetime}. Further, with the second-level communication group strategy, there will be no communication between supernodes. This ensures that our scaling result can be simply extended to large scale by submitting an individual task for each supernode. Our strategy successfully works on more than 1024 nodes (399360 cores). Combining all the optimization techniques, we can achieve more than 10 times speedup for Zuchongzhi-60-24 circuit \cite{zuchongzhi2}, which further pushes the boundaries of classic simulation.

\section{Implication}
In this work, we proposed a series of innovative TNC optimization for RQC simulation, and pushed the performance boundaries of TNC several steps further. With these methods, we can avoid the performance plunge when simulating large quantum circuits and achieve state-of-the-art efficiency. All the beneficial properties, such as parallelism and GEMM-implementation are still kept, while heavy memory access and huge computation overheads are largely reduced. Our multi-level optimizations covers memory access, communication and computation, offering up to 5 times speed-up compared with previous works. We believe the performance can be further improved by more fine-grained optimizations.

In addition to the progress we made in quantum circuit simulation, we are very concerned about the areas in which the proposed techniques will have impacts. Though our methods are implemented on Sunway, the parallel designs and ideas are not limited to certain architecture. Our core-array cooperation fusion and reuse design presents a new application scenario of multilevel parallel. Sometimes, embarrassing parallel is not the terminus of parallel design. Flexible use of multi-core collaboration may bring new chances for optimization, by detecting the intrinsic correlation of data. At instruction level, SIMD are often criticized for lack of flexibility. High performance and difficulties of algorithm design are inherent contradictions of long vector registers, or even for CPUs, due to the restriction of the number of cores. However, in this work, we showed that SIMD can play to its strengths with careful design and shuffle support. We are looking forward to more innovative SIMD algorithms and compiler integration. Furthermore, as shuffling can be implicitly done by SIMT, our vectorized permutation can be easily transported to GPU, avoiding bank conflict in some traditional partition methods.

The significance of High performance TNC will continue to grow in the future. As quantum computing technology continues to advance, the classical simulation of RQCs will remain a crucial tool for understanding the limitations and capabilities of quantum devices, guiding the development of quantum algorithms, and ultimately paving the way for practical applications that leverage the power of quantum computing. The race between quantum circuits and supercomputer will go on, and continue to provide new ideas to both quantum computing and supercomputing communities. For scientific computing and AI computing, TNC optimization also provides insights for more efficient kernels, algorithms and even hardwares. Fused design and high performance permutation can be directly transferred to AI computation with dense Einstein summations.

An important lesson we learnt from this work is the potential of complex systems and large-scale computing. The new problems and new strategies of large-scale TNC proves the famous saying: "More is different". Special properties will emerge when the size of systems pass a critical point. This is why scientists do researches on complex systems and supercomputer, instead of treating large-scale computation simply as an engineering issue. 



\bibliographystyle{ieeetr}
\bibliography{main.bib}

\begin{thebibliography}{10}

\bibitem{Preskill_2018}
J.~Preskill, ``Quantum computing in the nisq era and beyond,'' {\em Quantum},
  vol.~2, p.~79, Aug 2018.

\bibitem{google-nature-2019}
F.~Arute, K.~Arya, R.~Babbush, D.~Bacon, J.~C. Bardin, R.~Barends, R.~Biswas,
  S.~Boixo, F.~G. Brandao, D.~A. Buell, {\em et~al.}, ``Quantum supremacy using
  a programmable superconducting processor,'' {\em Nature}, vol.~574, no.~7779,
  pp.~505--510, 2019.

\bibitem{morvan2023phase}
A.~Morvan, B.~Villalonga, X.~Mi, S.~Mandra, A.~Bengtsson, P.~Klimov, Z.~Chen,
  S.~Hong, C.~Erickson, I.~Drozdov, {\em et~al.}, ``Phase transition in random
  circuit sampling,'' {\em arXiv preprint arXiv:2304.11119}, 2023.

\bibitem{TOP500}
``Top500: https://www.top500.org/lists/top500/2023/11/,'' 2023.

\bibitem{wille2019ibm}
R.~Wille, R.~Van~Meter, and Y.~Naveh, ``Ibm’s qiskit tool chain: Working with
  and developing for real quantum computers,'' in {\em 2019 Design, Automation
  \& Test in Europe Conference \& Exhibition (DATE)}, pp.~1234--1240, IEEE,
  2019.

\bibitem{li2021sv}
A.~Li, B.~Fang, C.~Granade, G.~Prawiroatmodjo, B.~Heim, M.~Roetteler, and
  S.~Krishnamoorthy, ``Sv-sim: scalable pgas-based state vector simulation of
  quantum circuits,'' in {\em Proceedings of the International Conference for
  High Performance Computing, Networking, Storage and Analysis}, pp.~1--14,
  2021.

\bibitem{zuchongzhi}
Q.~Zhu, S.~Cao, F.~Chen, M.-C. Chen, X.~Chen, T.-H. Chung, H.~Deng, Y.~Du,
  D.~Fan, M.~Gong, C.~Guo, C.~Guo, S.~Guo, L.~Han, L.~Hong, H.-L. Huang, Y.-H.
  Huo, L.~Li, N.~Li, S.~Li, Y.~Li, F.~Liang, C.~Lin, J.~Lin, H.~Qian, D.~Qiao,
  H.~Rong, H.~Su, L.~Sun, L.~Wang, S.~Wang, D.~Wu, Y.~Wu, Y.~Xu, K.~Yan,
  W.~Yang, Y.~Yang, Y.~Ye, J.~Yin, C.~Ying, J.~Yu, C.~Zha, C.~Zhang, H.~Zhang,
  K.~Zhang, Y.~Zhang, H.~Zhao, Y.~Zhao, L.~Zhou, C.-Y. Lu, C.-Z. Peng, X.~Zhu,
  and J.-W. Pan, ``Quantum computational advantage via 60-qubit 24-cycle random
  circuit sampling,'' {\em Science Bulletin}, vol.~67, no.~3, pp.~240--245,
  2022.

\bibitem{chen2018classical}
J.~Chen, F.~Zhang, C.~Huang, M.~Newman, and Y.~Shi, ``Classical simulation of
  intermediate-size quantum circuits,'' {\em arXiv preprint arXiv:1805.01450},
  2018.

\bibitem{wang2019benchmarking}
Y.~E. Wang, G.-Y. Wei, and D.~Brooks, ``Benchmarking tpu, gpu, and cpu
  platforms for deep learning,'' {\em arXiv preprint arXiv:1907.10701}, 2019.

\bibitem{jouppi2017datacenter}
N.~P. Jouppi, C.~Young, N.~Patil, D.~Patterson, G.~Agrawal, R.~Bajwa, S.~Bates,
  S.~Bhatia, N.~Boden, A.~Borchers, {\em et~al.}, ``In-datacenter performance
  analysis of a tensor processing unit,'' in {\em Proceedings of the 44th
  annual international symposium on computer architecture}, pp.~1--12, 2017.

\bibitem{gray2021hyper}
J.~Gray and S.~Kourtis, ``Hyper-optimized tensor network contraction,'' {\em
  Quantum}, vol.~5, p.~410, 2021.

\bibitem{zuchongzhi2}
Y.~Wu, W.-S. Bao, S.~Cao, F.~Chen, M.-C. Chen, X.~Chen, T.-H. Chung, H.~Deng,
  Y.~Du, D.~Fan, M.~Gong, C.~Guo, C.~Guo, S.~Guo, L.~Han, L.~Hong, H.-L. Huang,
  Y.-H. Huo, L.~Li, N.~Li, S.~Li, Y.~Li, F.~Liang, C.~Lin, J.~Lin, H.~Qian,
  D.~Qiao, H.~Rong, H.~Su, L.~Sun, L.~Wang, S.~Wang, D.~Wu, Y.~Xu, K.~Yan,
  W.~Yang, Y.~Yang, Y.~Ye, J.~Yin, C.~Ying, J.~Yu, C.~Zha, C.~Zhang, H.~Zhang,
  K.~Zhang, Y.~Zhang, H.~Zhao, Y.~Zhao, L.~Zhou, Q.~Zhu, C.-Y. Lu, C.-Z. Peng,
  X.~Zhu, and J.-W. Pan, ``Strong quantum computational advantage using a
  superconducting quantum processor,'' {\em Phys. Rev. Lett.}, vol.~127,
  p.~180501, Oct 2021.

\bibitem{huang2020classical}
C.~Huang, F.~Zhang, M.~Newman, J.~Cai, X.~Gao, Z.~Tian, J.~Wu, H.~Xu, H.~Yu,
  B.~Yuan, {\em et~al.}, ``Classical simulation of quantum supremacy
  circuits,'' {\em arXiv preprint arXiv:2005.06787}, 2020.

\bibitem{liu2021closing}
Y.~Liu, X.~Liu, F.~Li, H.~Fu, Y.~Yang, J.~Song, P.~Zhao, Z.~Wang, D.~Peng,
  H.~Chen, {\em et~al.}, ``Closing the" quantum supremacy" gap: achieving
  real-time simulation of a random quantum circuit using a new sunway
  supercomputer,'' in {\em Proceedings of the International Conference for High
  Performance Computing, Networking, Storage and Analysis}, pp.~1--12, 2021.

\bibitem{pan2021simulating}
F.~Pan and P.~Zhang, ``Simulating the sycamore quantum supremacy circuits,''
  {\em arXiv preprint arXiv:2103.03074}, 2021.

\bibitem{lifetime}
Y.~Chen, Y.~Liu, X.~Shi, J.~Song, X.~Liu, L.~Gan, C.~Guo, H.~Fu, J.~Gao,
  D.~Chen, and G.~Yang, ``Lifetime-based optimization for simulating quantum
  circuits on a new sunway supercomputer,'' in {\em Proceedings of the 28th ACM
  SIGPLAN Annual Symposium on Principles and Practice of Parallel Programming},
  PPoPP '23, (New York, NY, USA), p.~148–159, Association for Computing
  Machinery, 2023.

\bibitem{li2021sw_qsim}
F.~Li, X.~Liu, Y.~Liu, P.~Zhao, Y.~Yang, H.~Shang, W.~Sun, Z.~Wang, E.~Dong,
  and D.~Chen, ``Sw\_qsim: a minimize-memory quantum simulator with
  high-performance on a new sunway supercomputer,'' in {\em Proceedings of the
  International Conference for High Performance Computing, Networking, Storage
  and Analysis}, pp.~1--13, 2021.

\bibitem{walker1996mpi}
D.~W. Walker and J.~J. Dongarra, ``Mpi: a standard message passing interface,''
  {\em Supercomputer}, vol.~12, pp.~56--68, 1996.

\bibitem{fu2016sunway}
H.~Fu, J.~Liao, J.~Yang, L.~Wang, Z.~Song, X.~Huang, C.~Yang, W.~Xue, F.~Liu,
  F.~Qiao, {\em et~al.}, ``The sunway taihulight supercomputer: system and
  applications,'' {\em Science China Information Sciences}, vol.~59, no.~7,
  pp.~1--16, 2016.

\bibitem{villalonga2020establishing}
B.~Villalonga, D.~Lyakh, S.~Boixo, H.~Neven, T.~S. Humble, R.~Biswas, E.~G.
  Rieffel, A.~Ho, and S.~Mandr{\`a}, ``Establishing the quantum supremacy
  frontier with a 281 pflop/s simulation,'' {\em Quantum Science and
  Technology}, vol.~5, no.~3, p.~034003, 2020.

\bibitem{gray2018quimb}
J.~Gray, ``quimb: a python library for quantum information and many-body
  calculations,'' {\em Journal of Open Source Software}, vol.~3, no.~29,
  p.~819, 2018.

\bibitem{zhang2019alibaba}
F.~Zhang, C.~Huang, M.~Newman, J.~Cai, H.~Yu, Z.~Tian, B.~Yuan, H.~Xu, J.~Wu,
  X.~Gao, {\em et~al.}, ``Alibaba cloud quantum development platform:
  Large-scale classical simulation of quantum circuits,'' {\em arXiv preprint
  arXiv:1907.11217}, 2019.

\bibitem{paszke2019pytorch}
A.~Paszke, S.~Gross, F.~Massa, A.~Lerer, J.~Bradbury, G.~Chanan, T.~Killeen,
  Z.~Lin, N.~Gimelshein, L.~Antiga, {\em et~al.}, ``Pytorch: An imperative
  style, high-performance deep learning library,'' {\em Advances in neural
  information processing systems}, vol.~32, 2019.

\bibitem{cuquantum}
T.~cuQuantum~development team, ``Nvidia cuquantum sdk,'' Nov. 2023.

\bibitem{anderson1999lapack}
E.~Anderson, Z.~Bai, C.~Bischof, L.~S. Blackford, J.~Demmel, J.~Dongarra,
  J.~Du~Croz, A.~Greenbaum, S.~Hammarling, A.~McKenney, {\em et~al.}, {\em
  LAPACK users' guide}.
\newblock SIAM, 1999.

\bibitem{cutensor}
``cutensor: Tensor linear algebra on nvidia gpus,'' 2024.

\bibitem{CUTLASS}
V.~Thakkar, P.~Ramani, C.~Cecka, A.~Shivam, H.~Lu, E.~Yan, J.~Kosaian,
  M.~Hoemmen, H.~Wu, A.~Kerr, M.~Nicely, D.~Merrill, D.~Blasig, F.~Qiao,
  P.~Majcher, P.~Springer, M.~Hohnerbach, J.~Wang, and M.~Gupta, ``{CUTLASS},''
  Jan. 2023.

\bibitem{springer2018design}
P.~Springer and P.~Bientinesi, ``Design of a high-performance gemm-like
  tensor--tensor multiplication,'' {\em ACM Transactions on Mathematical
  Software (TOMS)}, vol.~44, no.~3, pp.~1--29, 2018.

\bibitem{goto2008anatomy}
K.~Goto and R.~A. v.~d. Geijn, ``Anatomy of high-performance matrix
  multiplication,'' {\em ACM Transactions on Mathematical Software (TOMS)},
  vol.~34, no.~3, pp.~1--25, 2008.

\bibitem{kalachev2021multi}
G.~Kalachev, P.~Panteleev, and M.-H. Yung, ``Multi-tensor contraction for xeb
  verification of quantum circuits,'' {\em arXiv preprint arXiv:2108.05665},
  2021.

\bibitem{liu2024verifying}
Y.~Liu, Y.~Chen, C.~Guo, J.~Song, X.~Shi, L.~Gan, W.~Wu, W.~Wu, H.~Fu, X.~Liu,
  {\em et~al.}, ``Verifying quantum advantage experiments with multiple
  amplitude tensor network contraction,'' {\em Physical Review Letters},
  vol.~132, no.~3, p.~030601, 2024.

\bibitem{higham2002accuracy}
N.~J. Higham, {\em Accuracy and stability of numerical algorithms}.
\newblock SIAM, 2002.

\bibitem{higham2022mixed}
N.~J. Higham and T.~Mary, ``Mixed precision algorithms in numerical linear
  algebra,'' {\em Acta Numerica}, vol.~31, pp.~347--414, 2022.

\bibitem{xu2018taming}
S.~Xu, Y.~Xu, W.~Xue, X.~Shen, F.~Zheng, X.~Huang, and G.~Yang, ``Taming the"
  monster": Overcoming program optimization challenges on sw26010 through
  precise performance modeling,'' in {\em 2018 IEEE International Parallel and
  Distributed Processing Symposium (IPDPS)}, pp.~763--773, IEEE, 2018.

\end{thebibliography}

\vspace{12pt}

\end{document}